\documentclass[twoside]{article}
\usepackage{qic,epsfig}

\newcommand{\comment}[1]{}
\newcommand{\Upshape}{}

\newcommand{\delete}[1]{}

\def\>{\rangle}
\def\<{\langle}

\begin{document}
\setlength{\textheight}{8.0truein}    

\runninghead{Title  $\ldots$}
            {Author(s) $\ldots$}

\normalsize\textlineskip
\thispagestyle{empty}
\setcounter{page}{1}

\copyrightheading{0}{0}{2010}{000--000}

\vspace*{0.88truein}

\alphfootnote

\fpage{1}

\centerline{\bf
Optimal Perfect Distinguishability between Unitaries and Quantum Operations}
\vspace*{0.035truein}
\vspace*{0.37truein}
\centerline{\footnotesize
CHENG LU \footnote{Email: {lvc.thu@gmail.com}.}}
\vspace*{0.015truein}
\centerline{\footnotesize\it Department of Computer Science and Technology, Tsinghua University}
\baselineskip=10pt
\centerline{\footnotesize\it Beijing 100084, CHINA}
\vspace*{10pt}

\centerline{\footnotesize
JIANXIN CHEN \footnote{Email: {chenkenshin@gmail.com}.}}
\vspace*{0.015truein}
\centerline{\footnotesize\it State Key Laboratory of Intelligent Technology and Systems,}
\centerline{\footnotesize\it Department of Computer Science and Technology,}
\centerline{\footnotesize\it Tsinghua National Laboratory for Information Science and Technology,}
\baselineskip=10pt
\centerline{\footnotesize\it Tsinghua University, Beijing 100084, China}
\vspace*{10pt}

\centerline{\footnotesize
RUNYAO DUAN\footnote{Email: {runyao.duan@uts.edu.au}.
Supported in part by QCIS of University of Technology, Sydney, and the National Natural Science Foundation of China
(Grant Nos. 60702080 and 60621062).}}
\vspace*{0.015truein}
\centerline{\footnotesize\it Centre for Quantum Computation and Intelligent Systems (QCIS),\\
}
\centerline{\footnotesize\it Faculty of Engineering and Information Technology,\\
}
\centerline{\footnotesize\it University of Technology, Sydney (UTS), NSW 2007, Australia \\
}

\vspace*{0.225truein}
\publisher{}{}

\vspace*{0.21truein}

\abstracts{
We study optimal perfect distinguishability between a unitary and a general quantum operation. In $2$-dimensional case we provide a simple sufficient and necessary condition for sequential perfect distinguishability and an analytical formula of optimal query time. We extend the sequential condition to general $d$-dimensional case. Meanwhile, we provide an upper bound and a lower bound for optimal sequential query time. In the process a new iterative method is given, the most notable innovation of which is its independence to auxiliary systems or entanglement. Following the idea, we further obtain an upper bound and a lower bound of (entanglement-assisted) $q$-maximal fidelities between a unitary and a quantum operation. Thus by the recursion in \cite{DFY09} an upper bound and a lower bound for optimal general perfect discrimination are achieved. Finally our lower bound result can be extended to the case of arbitrary two quantum operations.}{}{}

\vspace*{10pt}

\keywords{quantum operations, perfect distinguishability, maximal fidelity, $q$-maximal fidelity, optimal query time}
\vspace*{3pt}
\communicate{to be filled by the Editorial}

\vspace*{1pt}\textlineskip    

\newcommand{\EPR}{|\Psi^-\rangle}
\newcommand{\bra}[1]{\langle #1|}
\newcommand{\ket}[1]{|#1\rangle}
\newcommand{\braket}[3]{\langle #1|#2|#3\rangle}
\newcommand{\ip}[2]{\langle #1|#2\rangle}
\newcommand{\op}[2]{|#1\rangle \langle #2|}

\newcommand{\tr}{{\rm tr}}
\newcommand{\supp}{{\rm supp}}
\newcommand{\sch}{{\rm Sch}}

\newcommand{\Span}{\mathrm{span}}
\newcommand {\E } {{\mathcal{E}}}
\newcommand{\In}{\mathrm{in}}
\newcommand{\Out}{\mathrm{out}}
\newcommand{\local}{\mathrm{local}}
\newcommand {\F } {{\mathcal{F}}}
\newcommand {\diag } {{\rm diag}}
\renewcommand{\b}{\mathcal{B}}
\newcommand{\h}{\mathcal{H}}
\newcommand{\HSPACE}{}
\renewcommand{\Re}{\mathrm{Re}}
\renewcommand{\Im}{\mathrm{Im}}
\newcommand{\Z}{\sigma_z}
\newcommand{\Clocal}{C^{(0)}_{\local}}
\newcommand{\Sp}[1][p]{S_{\min}^{(#1)}}
\newcommand {\M} {{\mathcal{M}}}
\newcommand {\f } {\tilde{F}}
\newcommand {\R } {{\mathcal{R}}}
\newcommand {\I } {{\mathcal{I}}}
\newcommand{\T}{\mathcal{T}}

\newtheorem{proposition}[theorem]{Proposition}
\section{Introduction}
Perfect distinguishability of quantum operations is a fundamental problem
in quantum information theory. Informally, an unknown quantum operation is selected secretly from a set of known quantum operations, the problem is to identify the unknown quantum operation perfectly by finite queries to the operation with possible auxiliary systems. This problem has not only strong physical application background, but also impacts to other important problems in quantum information theory, such as superdence coding, oracle based quantum search algorithms, etc. Although perfect distinguishability between quantum states is completely characterized by orthogonality, perfect distinguishability of quantum operations is sharply different\cite{DFY09}. Consequently, the problem arouses people's great interests and a series of results are obtained\cite{DFY09,ACMB+07,AK97,AKN97,CP00,AC01,DLP01,SAC05,DY073,WY06,AC07,JW08,MZ08,PW09,DFY07,WD08,CDP08,CY10,JFDY06}, especially \cite{CP00,AC01,DLP01,DFY07,WD08,CDP08,CY10} solve perfect distinguishability between unitaries and \cite{JFDY06} solves perfect distinguishability between projective measurements.

Most notably, feasible sufficient and necessary conditions for perfect distinguishability between general quantum operations are completely obtained in \cite{DFY09}, where an algorithm for optimal perfect discrimination is given. However, although it is possible to estimate the optimal query time with arbitrary high precision based on the iteration\cite{DFY09}, computation is quite complex. Therefore for some special pairs of quantum operations, it is urgent to further quantify the optimal query time for perfect discrimination, which would also provide guidance to physical experiments.

In this letter, we study optimal perfect distinguishability between a unitary and a quantum operation. For convenience, we first give a uniform introduction to the notations. In a $d$-dimensional Hilbert space $\h_d$, the set of linear
operators on $\h_d$ is denoted by $\b(\h_d)$. A quantum density operator (quantum state) $\rho$ on $\h_d$ is given by a positive operator in $\b(\h_d)$
with trace one. A pure state $\ket{\psi} \in \h_d$ is a unit vector. For simplicity, we shall use $\psi$ to denote the density operator form $\op{\psi}{\psi}$ of $\ket{\psi}$. Since $\rho$ is positive, it has spectral decomposition $\rho=\sum_{i=1}^d p_i\op{\psi_i}{\psi_i}$ where $\{\ket{\psi_i}\}_{i=1}^d$ are mutually orthonormal vectors. The support of $\rho$ is defined as ${\rm supp}(\rho)={\rm span}\{\ket{\psi_i}:p_i>0\}$. Most importantly, a $d$-dimensional quantum operation $\E$ from $\b(\h_d)$ to $\b(\h_{d'})$ is a trace-preserving completely positive map with the form $\E(\rho)=\sum_{i=1}^m E_i\rho E_i^\dagger$, where $\{E_i\}_{i=1\cdots m}$ are the Kraus operators of $\E$ satisfying $\sum_{i}E_i^\dagger E_i=I_d$. Finally, it is easy to verify that for any pure states $\psi = \op{\psi}{\psi}$, ${\rm supp}(\E(\psi))={\rm span}\{E_i\ket{\psi}\}_{i=1\cdots m}$.

The maximal fidelity between two quantum states $\rho_0$ and $\rho_1$ is defined as $\f(\rho_0,\rho_1) = \max\{|\ip{\psi_0}{\psi_1}|: \ket{\psi_k} \in {\rm supp}(\rho_k),k=0,1\}$. The $q$-maximal fidelity between two quantum operations $\E_0$ and $\E_1$ with $0 \leq q \leq 1$ is defined as $\f_q(\E_0,\E_1) = \min\{\f(\E_0(\psi_0),\E_1(\psi_1)): \ip{\psi_0}{\psi_1}=q\}$. The entanglement-assisted $q$-maximal fidelity is defined as $\f_q^{ea}(\E_0,\E_1)=\f_q(\I^R \otimes \E_0^Q,\I^R \otimes \E_1^Q)$, where $R$ is an auxiliary system with the same dimension as $Q$ (larger dimensions make no difference)\cite{DFY09}.

\section{Sequential Optimal Perfect Distinguishability between a Unitary and a Quantum Operation}
\noindent
It is worth noting that optimal perfect distinguishability between a unitary and a quantum operation is equivalent to that between an identity and a quantum operation. Given an unknown operation $X$ selected from unitary $U$ and $\E$, by appending operation $U^\dagger$, the problem is reduced to optimal perfect distinguishability between $U^\dagger U=\I$ and $U^\dagger\E =\E'$. Conversely, as $\I$ is a unitary, it is trivial. In view of this, we shall only consider optimal perfect distinguishability between an identity and a quantum operation. We shall start from the sequential cases. Note that in sequential schemes no auxiliary systems or entanglement are allowed.

\begin{theorem}\Upshape\label{seq2dim}
$\E=\sum_{i=1}^m E_i \cdot E_i^\dagger(2 \leq m \leq 3)$ and $\I$ are 2-dimensional quantum operation and identity respectively. $\E$ and $\I$ are sequentially perfect distinguishable iff $\f_1(\E,\I) < 1$. The minimum query time is $N_{\min} = \biggl\lceil\frac{\pi}{2\arccos(\f_1(\E,\I))}\biggr\rceil$.
\end{theorem}

{\bf Proof:} The necessity can be proved by contradiction. Suppose $\f_1(\E, \I) = 1$ and the optimal query time is $N_{\min}$ with initial input state $\ket{b}$. According to the definition of $\f_1$, $\ket{b} \in {\rm supp}(\E(b))$, which means that the rest of the $N_{\min}-1$ queries can also complete discrimination by inputing $\ket{b}$. It contradicts the minimality of $N_{\min}$.

When $\f_1(\E,\I)<1$, let $\f_1(\E,\I)=\cos\theta(0 < \theta \leq \frac{\pi}{2}), \ket{b}: \f(\E(b),b)=\f_1(\E,\I)$. In $2$-dimensional case, it is equivalent to $\E(b)=c$ and $|\ip{b}{c}|=\cos\theta$. Take $\ket{b}$ as the initial input state. W.l.o.g, let $\ket{c}=\cos\theta\ket{b}+\sin\theta\ket{b^\perp}$.

If $\f_1(\E,\I)=0$, $\E$ and $\I$ can transform $\ket{b}$ into two orthogonal pure states. Then one projective measurement is enough to distinguish the two states, thus distinguishing the two operations perfectly. Otherwise, most generally, before the next query any known intermediate quantum operations are acceptable. Let the operation be $\T_1: \T_1(c) = c_1$, $\T_1(b) = b_1$ and $\cos\theta=|\ip{c}{b}|\leq |\ip{c_1}{b_1}| \leq 1$\cite{DFY09}.
\begin{enumerate}
\item $2\theta < \frac{\pi}{2}$: get $\T_1$ such that $\ket{c_1}=\ket{b}$, $\ket{b_1} = \cos\theta\ket{b}-\sin\theta\ket{b^\perp}$;
\item $2\theta \geq \frac{\pi}{2}$: get $\T_1$ such that $\ket{c_1}=\ket{b}$, $\ket{b_1}=\sin\theta\ket{b}-\cos\theta\ket{b^\perp}$.
\end{enumerate}
Take $\ket{c_1}$ and $\ket{b_1}$ as respective inputs to the second query.

Afterward, if $2\theta \geq \frac{\pi}{2}$, $|\ip{c}{b_1}|=0$. Similarly discrimination completes. Otherwise $2\theta < \frac{\pi}{2}$, $|\ip{c}{b_1}|=\cos(2\theta)$. Further query is needed.

Inductively, suppose after querying the unknown operator $k(k \geq 2)$ times we obtain respective output states $\ket{c}$ and $\ket{b_{k-1}}$ for $\E$ and $\I$. If $k\theta \geq \frac{\pi}{2}$, $|\ip{c}{b_{k-1}}| = 0$, we can perfectly distinguish the two operators. Otherwise, $k\theta < \frac{\pi}{2}$, $|\ip{c}{b_{k-1}}| = \cos(k\theta)$, we can't still complete perfect discrimination. If so, we shall prove that after $k+1$ querying times(one more query based on the previous $k$ queries), we can obtain respective output states to $\E$ and $\I$ as $\ket{c}$ and $\ket{b_k}$ such that if $(k+1)\theta \geq \frac{\pi}{2}$, $|\ip{c}{b_k}|=0$, otherwise $|\ip{c}{b_k}| = \cos((k+1)\theta)$, similar to the case of $k$. Note that the cases where $k=1$ and $k=2$ have been proved above, where $\ket{b_0}=\ket{b}$.

To make it most general, before the $(k+1)$th query, an operation $\T_k$ is used beforehand. Let $\T_k(c)=c_k, \T_k(b_{k-1})=b_k$, $\cos(k\theta)=|\ip{c}{b_{k-1}}| \leq |\ip{c_k}{b_k}| \leq 1$\cite{DFY09}.
\begin{enumerate}
\item $(k+1)\theta < \frac{\pi}{2}$: get $\T_k$ such that $\ket{c_k} = \ket{b}$, $\ket{b_k} = \cos(k\theta)\ket{b}-\sin(k\theta)\ket{b^\perp}$;
\item $(k+1)\theta \geq \frac{\pi}{2}$: get $\T_k$ such that $\ket{c_k}=\ket{b}$, $\ket{b_k} = \sin\theta\ket{b}-\cos\theta\ket{b^\perp}$.
\end{enumerate}
Take $\ket{c_k}$ and $\ket{b_k}$ as respective inputs to the $(k+1)$th query. Afterward, if $(k+1)\theta \geq \frac{\pi}{2}$, $|\ip{c}{b_k}|=0$ and it is done. Otherwise $(k+1)\theta < \frac{\pi}{2}$, $|\ip{c}{b_k}|=\cos((k+1)\theta)$. Hence we should repeat again for the next query. Therefore we prove that the conclusions also hold for the case of $k+1$.

By induction, we prove that the above conclusions hold for every integer $k \geq 1$. Since $0 < \theta \leq \frac{\pi}{2}$ is a fixed number, finally we can definitely obtain two states $c$ and $b_{n-1}$ from $\E$ and $\I$ respectively such that $|\ip{c}{b_{n-1}}|=0$ and complete discrimination. The above scheme requires
\begin{equation}
\biggl\lceil\frac{\pi}{2\arccos(\f_1(\E,\I))}\biggr\rceil
\end{equation}
queries. Now we prove that it is optimal.

In the optimal sequential scheme, the initial state $\ket{b}$ should satisfy $\f_1(\E,\I)\leq \f(\E(b),b)<1$. $2$-dimensionally, $\E(b)=c$ and $|\ip{c}{b}|=\f(\E(b),b)$. Let $\f_1(\E,\I)=\cos\theta(0 < \theta \leq \frac{\pi}{2}), |\ip{c}{b}|=\cos\psi(0 \leq \psi \leq \frac{\pi}{2})$, hence $0 < \psi \leq \theta$. If $\theta = \frac{\pi}{2}$, get $\psi = \theta = \frac{\pi}{2}$, then one query can complete the task. It coincides with the above scheme.

Otherwise, at least the second query is needed. Beforehand, assume $\T_1$ is used such that $\T_1(c)=c_1, \T_1(b)=b_1$ and $\cos\psi=|\ip{c}{b}| \leq |\ip{c_1}{b_1}| = \cos\beta_1<1$\cite{DFY09}, or $0 < \beta_1 \leq \psi \leq \theta$. After the second query, $\E$ should output a pure state, otherwise in $2$-dimensional case, there must be a common state belonging to both supports of the output states, thus breaking optimality. Let $\ket{c'_1}$ and $\ket{b_1}$ be the respective output states of $\E$ and $\I$.
\begin{eqnarray}
|\ip{c_1'}{b_1}| &=& |\ip{c_1'}{c_1}\ip{c_1}{b_1}+\ip{c_1'}{c_1^\perp}\ip{c_1^\perp}{b_1}|\nonumber \\
    &\geq& \big||\ip{c_1'}{c_1}\ip{c_1}{b_1}|-|\ip{c_1'}{c_1^\perp}\ip{c_1^\perp}{b_1}|\big|\nonumber \\
    &=& |\cos\psi_1\cos\beta_1-\sin\psi_1\sin\beta_1|\nonumber \\
    &=& |\cos(\psi_1+\beta_1)|,
\end{eqnarray}
where $\cos\theta = \f_1(\E,\I) \leq \cos\psi_1 = |\ip{c_1'}{c_1}| < 1$, or $0 < \psi_1 \leq \theta$. If $2\theta \geq \frac{\pi}{2}$, $|\ip{c'_1}{b_1}| \geq 0$, 0 is possible to reach. Meanwhile the above scheme can right obtain 0, hence it is optimal. If $2\theta < \frac{\pi}{2}$,
\begin{equation}
|\ip{c_1'}{b_1}| \geq |\cos(\psi_1+\beta_1)| \geq |\cos(2\theta)| = \cos(2\theta) > 0.
\end{equation}
The third query is needed.

Inductively, in the optimal scheme, suppose after querying the unknown operator $k(k \geq 2)$ times we obtain respective output states $\ket{c'_{k-1}}$ and $\ket{b_{k-1}}$ for $\E$ and $\I$. If $k\theta \geq \frac{\pi}{2}$, $|\ip{c'_{k-1}}{b_{k-1}}| \geq 0$, we can possibly perfectly distinguish the two operators. Otherwise, $k\theta < \frac{\pi}{2}$, $|\ip{c'_{k-1}}{b_{k-1}}| \geq \cos(k\theta)>0$, we can't still complete perfect discrimination. If so, we shall prove that after $k+1$ querying times(one more query based on the previous $k$ queries), we can obtain respective output states to $\E$ and $\I$ as $\ket{c'_k}$ and $\ket{b_k}$ such that if $(k+1)\theta \geq \frac{\pi}{2}$, $|\ip{c'_k}{b_k}| \geq 0$, otherwise $|\ip{c'_k}{b_k}| \geq \cos((k+1)\theta)>0$, similar to the case of $k$. Note that the cases where $k=1$ and $k=2$ have been proved above, where $\ket{c'_0}=\ket{c}$ and $\ket{b_0}=\ket{b}$.

To make it most general, before the $(k+1)$th query, assume the intermediate operation is $\T_k$, $\T_k(c'_{k-1})=c_k, \T_k(b_{k-1})=b_k$\cite{DFY09}. Let $\E(c_k)=c'_k$. We can obtain a similar lower bound:
\begin{eqnarray}
|\ip{c_k'}{b_k}| &=& |\ip{c_k'}{c_k}\ip{c_k}{b_k}+\ip{c_k'}{c_k^\perp}\ip{c_k^\perp}{b_k}|\nonumber \\
    &\geq& \big||\ip{c_k'}{c_k}\ip{c_k}{b_k}|-|\ip{c_k'}{c_k^\perp}\ip{c_k^\perp}{b_k}|\big|\nonumber \\
    &=& |\cos\psi_k\cos\beta_k-\sin\psi_k\sin\beta_k|\nonumber \\
    &=& |\cos(\psi_k+\beta_k)|,
\end{eqnarray}
where $\cos\psi_k=|\ip{c'_k}{c_k}|, \cos\beta_k=|\ip{c_k}{b_k}|(0 \leq \psi_k,\beta_k \leq \frac{\pi}{2})$. Hence $\cos\theta=\f_1(\E,\I) \leq |\ip{c'_k}{c_k}|=\cos\psi_k < 1$ and $\cos(k\theta) \leq |\ip{c'_{k-1}}{b_{k-1}}| \leq |\ip{c_k}{b_k}| = \cos\beta_k < 1$\cite{DFY09}, or $0 < \psi_k \leq \theta$ and $0 < \beta_k \leq k\theta$. If $(k+1)\theta \geq \frac{\pi}{2}$, $|\ip{c'_k}{b_k}| \geq 0$, 0 is possible to reach. The above scheme is shown optimal. Otherwise,
\begin{equation}
|\ip{c_k'}{b_k}| \geq |\cos(\psi_k+\beta_k)| \geq |\cos((k+1)\theta)| = \cos((k+1)\theta) > 0.
\end{equation}
Hence we should repeat again for the next query. Therefore we prove that the conclusions also hold for the case of $k+1$.

By induction, we prove that the above lower bounds hold for every integer $k \geq 1$ and the above given scheme is optimal. Therefore in $2$-dimensional case the minimum query time is
\begin{equation}
    N_{\min} = \biggl\lceil\frac{\pi}{2\arccos(\f_1(\E,\I))}\biggr\rceil.
\end{equation}
\hfill $\square$

In Theorem \ref{seq2dim}, when $m=1$, the conclusion matches with that in \cite{DFY07} perfectly. The above proof is enlightening. It is natural to consider extension to arbitrary dimensions. Interestingly, the answer is partially positive. First we need the following lemma.

\begin{lemma}\label{upperbound}\Upshape
$\E = \sum_{i=1}^m E_i \cdot E_i^\dagger (2 \leq m \leq d^2-1)$ and $\I$ are $d$-dimensional quantum operation and identity respectively. $q = \cos\alpha(0 \leq \alpha \leq \frac{\pi}{2}), \f_1(\E,\I)=\cos\theta(0 < \theta < \frac{\pi}{2})$, then
\begin{equation}
\f_q(\E,\I) \leq \frac{|\sin(\alpha_0-\alpha)|}{\sin\alpha_0}\cos\theta,
\end{equation}
where $\cos\alpha_0 = \max\{|\ip{b}{b'}|: \f(\E(b),b)=\cos\theta, \op{b}{b'} \in {\rm span}^\perp \{E_i\}_{i = 1\ldots m}\}(0 \leq \alpha_0 \leq \frac{\pi}{2})$.
\end{lemma}

{\bf Proof:} First we shall prove $\alpha_0$ exists and $0 < \alpha_0 < \frac{\pi}{2}$. Since $\theta > 0$, $\exists \ket{b}$ such that $\f(\E(b),b)=\f_1(\E,\I)=\cos\theta < 1$ and $\ket{b} \notin {\rm supp}(\E(b))$. ${\rm dim}({\rm supp}(\E(b))) \leq d-1$. Equivalently, ${\rm dim}({\rm supp}^\perp(\E(b))) \geq 1$. Hence $\exists \ket{b'} \in {\rm supp}^\perp(\E(b))$, or $\op{b}{b'} \in {\rm span}^\perp\{E_i\}_{i=1\ldots m}$. Thus $\alpha_0$ exists. What's more, since $\theta < \frac{\pi}{2}$, $\alpha_0 > 0$. By $\f(\E(b),b)<1$, $\alpha_0 < \frac{\pi}{2}$. Get $\ket{b'}$ such that $\ip{b}{b'}=\cos\alpha_0$.

Let $\ket{b}$ and $\ket{c}$ are input states of $\E$ and $\I$ respectively, $\ip{b}{c}=q=\cos\alpha$ and $\ket{c}=x\ket{b}+y\ket{b'}(x,y \in \R, y > 0)$, where $x$ and $y$ are to be determined. Note that since $\ip{b}{b'}=\cos\alpha_0<1$, such $\ket{c}$ exists. By $\ip{b}{c}=q=\cos\alpha$ and $\ip{c}{c}=1$, we have
\begin{eqnarray}
    x + y\cos\alpha_0 = \cos\alpha, \\
    x^2 + 2xy\cos\alpha_0 + y^2 = 1.
\end{eqnarray}
Hence
\begin{eqnarray}
    y &=& \sqrt{\frac{1-\cos^2\alpha}{1-\cos^2\alpha_0}} = \frac{\sin\alpha}{\sin\alpha_0}, \\
    x &=& \cos\alpha - y\cos\alpha_0 = \frac{\sin(\alpha_0-\alpha)}{\sin\alpha_0}.
\end{eqnarray}
For $\forall \ket{d} \in {\rm supp}(\E(b))$,
\begin{eqnarray}
    |\ip{d}{c}| &=& |x\ip{d}{b}+y\ip{d}{b'}| = |x\ip{d}{b}|\nonumber \\
            &\leq& |x|\f(\E(b),b) = \frac{|\sin(\alpha_0-\alpha)|}{\sin\alpha_0}\cos\theta.
\end{eqnarray}
By the definition of $\f_q(\E,\I)$, finally we have
\begin{equation}
    \f_q(\E,\I) \leq \f(\E(b),c) \leq \frac{|\sin(\alpha_0-\alpha)|}{\sin\alpha_0}\cos\theta.
\end{equation}
\hfill $\square$

\begin{theorem}\Upshape\label{seqddim}
$\E = \sum_{i=1}^m E_i \cdot E_i^\dagger (2 \leq m \leq d^2-1)$ and $\I$ are $d$-dimensional quantum operation and identity respectively. $\E$ and $\I$ are sequentially perfectly distinguishable iff $\f_1(\E,\I)=\cos\theta < 1$. The minimum query time $N_{\min}$ satisfies
\begin{equation}
    \biggl\lceil\frac{\pi}{2\arccos(\f_1(\E,\I))}\biggr\rceil \leq N_{\min} \leq \biggl\lceil\frac{\ln \cos\alpha_0}{\ln \cos\theta}\biggr\rceil+1,
\end{equation}
where $\cos\alpha_0 = \max\{|\ip{b}{b'}|: \f(\E(b),b)=\cos\theta, \op{b}{b'} \in {\rm span}^\perp \{E_i\}_{i = 1\ldots m}\}(0 \leq \alpha_0 \leq \frac{\pi}{2})$ and define $\ln 0=1$.
\end{theorem}

{\bf Proof:} The necessity part is the same as that in Theorem \ref{seq2dim} by contradiction. For sufficiency, suppose $\ket{b}$ satisfies $\f(\E(b),b)=\f_1(\E,\I)=\cos\theta < 1$. If $\f_1(\E,\I)=0$, one query with input $\ket{b}$ can complete the job.

Otherwise $0 < \theta < \frac{\pi}{2}$. $\forall 0 \leq \alpha \leq \alpha_0({\rm or }\cos\alpha_0 \leq \cos\alpha \leq 1)$, we have
\begin{equation}
    \frac{|\sin(\alpha_0-\alpha)|}{\sin\alpha_0} \leq \cos\alpha.
\end{equation}
Hence according to Lemma \ref{upperbound},
\begin{equation}
    \f_{q=\cos\alpha}(\E,\I) \leq \frac{|\sin(\alpha_0-\alpha)|}{\sin\alpha_0}\cos\theta \leq \cos\alpha\cos\theta.
\end{equation}
In other words, when $\cos\alpha_0 \leq q = \cos\alpha \leq 1$, every one query can reduce current maximal fidelity $q$ by at least a factor of $\cos\theta$.

Therefore we have the following scheme. Input $\ket{b}$ initially such that $\f(\E(b),b)=\f_1(\E,\I)=\cos\theta = q_1$. Apply an operation $\T_1$ to transform $\{\E(b),b\}$ to $\{\ket{c_1}, \ket{b_1}\}$ respectively such that $\ip{c_1}{b_1}=q_1$\cite{DFY09} and $\f(\E(c_1),b_1)=\f_{q_1}(\E,\I)=q_2$. Based on the above analysis, $q_2 \leq q_1\cos\theta < q_1$. Inductively, before the $k$th$(k \geq 2)$ query to the unknown operator, we use an operation $\T_{k-1}$ to transform $\{\E(c_{k-2}), b_{k-2}\}(\ket{c_0}=\ket{b_0}=\ket{b})$ into $\{\ket{c_{k-1}},\ket{b_{k-1}}\}$ respectively such that $\ip{c_{k-1}}{b_{k-1}}=q_{k-1}$\cite{DFY09} and $\f(\E(c_{k-1}),b_{k-1})=\f_{q_{k-1}}(\E,\I)=q_k$, where $q_k \leq q_{k-1}\cos\theta < q_{k-1}$. In this way, finally we can reach some $q_n \leq \cos\alpha_0$. Use another operation $\T_n$ to transform current states $\{\E(c_{n-1}),b_{n-1}\}$ to $\{\ket{b},\ket{b'}\}$ respectively\cite{DFY09}. By inputting $\ket{b}$ and $\ket{b'}$ to $\E$ and $\I$ respectively for one more query, we can obtain two orthogonal states. Finally a projective measurement can complete perfect discrimination. The scheme takes $\biggl\lceil\frac{\ln \cos\alpha_0}{\ln \cos\theta}\biggr\rceil+1$ queries. Therefore $N_{\min} \leq \biggl\lceil\frac{\ln \cos\alpha_0}{\ln \cos\theta}\biggr\rceil+1$.

$N_{\min}$'s lower bound: consider the optimal sequential scheme. Initial input $\ket{b}$ should satisfy $\f_1(\E,\I) \leq \f(\E(b),b) < 1$. Assume $\ket{c} \in {\rm supp}(\E(b))$ and $|\ip{c}{b}|=\f(\E(b),b)=\cos\psi(0 < \psi \leq \theta)$. If $\theta = \frac{\pi}{2}$, get $\psi = \theta = \frac{\pi}{2}$, then one query is enough for perfect discrimination. Here $N_{\min} = 1 \geq \biggl\lceil\frac{\pi}{2\arccos(\f_1(\E,\I))}\biggr\rceil = 1$. Lower bound holds.

Otherwise, $0 < \theta < \frac{\pi}{2}$, further queries are needed. Beforehand, suppose the operation is $\T_1$, $\T_1(\E(b))=c_1, \T_1(b)=b_1$ such that $\cos\psi = |\ip{c}{b}| \leq |\ip{c_1}{b_1}|=\cos\beta_1 < 1$\cite{DFY09}, or $0 < \beta_1 \leq \psi \leq \theta$. After the second query, let $t_1 = \f(\E(c_1),b_1)$. A key property is that $\exists \ket{c_1'} \in {\rm supp}(\E(c_1))$ such that $|\ip{c_1'}{c_1}| = \f(\E(c_1),c_1)=\cos\psi_1$ and $\ket{c_1'} = \cos\psi_1\ket{c_1}+\sin\psi_1\ket{c_1^\perp}$, where $\ip{c_1^\perp}{c_1}=0$. Assume $\{\ket{c_1},\ket{c_1^\perp},\ket{c_{13}}, \ldots, \ket{c_{1d}}\}$ form an orthonormal basis on $d$-dimensional space, and $\ket{b_1} = \cos\beta_1\ket{c_1} + e^{i\eta_1}\sin\gamma_1\ket{c_1^\perp}+\ket{b_1'}$, where $\ket{b_1'} \in {\rm span}\{\ket{c_{13}}, \ldots, \ket{c_{1d}}\}$ may not be normalized, $0 \leq \sin\gamma_1 \leq \sin\beta_1$. Thus
\begin{eqnarray}
    t_1 \geq |\ip{c_1'}{b_1}| &=& |\cos\psi_1\cos\beta_1+e^{i\eta_1}\sin\psi_1\sin\gamma_1|\nonumber \\
    &\geq& |\cos\psi_1\cos\beta_1-\sin\psi_1\sin\gamma_1|.
\end{eqnarray}
Since $0 \leq \sin\gamma_1 \leq \sin\beta_1$, $\cos(\psi_1+\beta_1) \leq \cos\psi_1\cos\beta_1-\sin\psi_1\sin\gamma_1 \leq \cos\psi_1\cos\beta_1$. On the other hand, since $\cos\theta \leq \cos\psi_1 < 1$ and $\cos\theta \leq \cos\psi \leq \cos\beta_1 < 1$, we have $\cos(\psi_1+\beta_1) \geq \cos(2\theta)$.

If $2\theta \geq \frac{\pi}{2}$, $|\cos\psi_1\cos\beta_1-\sin\psi_1\sin\gamma_1| \geq 0$ and it is possible that $t_1 = \f(\E(c_1),b_1) = 0$. If so, finally one projective measurement can complete the task. Then we have $N_{\min} \geq \biggl\lceil\frac{\pi}{2\arccos(\f_1(\E,\I))}\biggr\rceil$. Otherwise, $2\theta < \frac{\pi}{2}$, $t_1 \geq |\cos\psi_1\cos\beta_1-\sin\psi_1\sin\gamma_1| \geq \cos(2\theta) > 0$. Further queries are necessary.

Inductively, in the optimal scheme, suppose after querying the unknown operator $k(k \geq 2)$ times we obtain respective output states $\E(c_{k-1})$ and $b_{k-1}$ for $\E$ and $\I$. If $k\theta \geq \frac{\pi}{2}$, $t_{k-1} = \f(\E(c_{k-1}),b_{k-1}) \geq 0$, we can possibly perfectly distinguish the two operators and the lower bound holds. Otherwise, $k\theta < \frac{\pi}{2}$, $t_{k-1} = \f(\E(c_{k-1}),b_{k-1}) \geq \cos(k\theta) > 0$, we can't still complete perfect discrimination. If so, we shall prove that after $k+1$ querying times(one more query based on the previous $k$ queries), we can obtain respective output states to $\E$ and $\I$ as $\E(c_k)$ and $b_k$ such that if $(k+1)\theta \geq \frac{\pi}{2}$, $t_k = \f(\E(c_k),b_k) \geq 0$, we can possibly perfectly distinguish the two operators and the lower bound holds. Otherwise $t_k = \f(\E(c_k),b_k) \geq \cos((k+1)\theta) > 0$, similar to the case of $k$. Note that the cases where $k=1$ and $k=2$ have been proved above, where $\ket{c_0}=\ket{b_0}=\ket{b}$.

Before the $(k+1)$th query, we apply an operation $\T_k$, $\T_k(\E(c_{k-1}))=c_k, \T_k(b_{k-1})=b_k$, and $\cos(k\theta) \leq t_{k-1} \leq |\ip{c_k}{b_k}|=\cos\beta_k<1$\cite{DFY09}, or $0 < \beta_k \leq k\theta$. After the $(k+1)$th query, let $t_k=\f(\E(c_k),b_k)$. The same property also holds that $\exists \ket{c_k'} \in {\rm supp}(\E(c_k))$ such that $|\ip{c_k'}{c_k}|=\f(\E(c_k),c_k)=\cos\psi_k$ and $\ket{c_k'} = \cos\psi_k\ket{c_k}+\sin\psi_k\ket{c_k^\perp}$, where $\ip{c_k^\perp}{c_k}=0$. Suppose $\{\ket{c_k},\ket{c_k^\perp},\ket{c_{k3}}, \ldots, \ket{c_{kd}}\}$ form an orthonormal basis over $d$-dimensional space, and $\ket{b_k} = \cos\beta_k\ket{c_k} + e^{i\eta_k}\sin\gamma_k\ket{c_k^\perp}+\ket{b_k'}$, where $\ket{b_k'} \in {\rm span}\{\ket{c_{k3}}, \ldots, \ket{c_{kd}}\}$ may not be normalized, and $0 \leq \sin\gamma_k \leq \sin\beta_k$. We can obtain a similar lower bound:
\begin{eqnarray}
    t_k \geq |\ip{c_k'}{b_k}| &=& |\cos\psi_k\cos\beta_k+e^{i\eta_k}\sin\psi_k\sin\gamma_k|\nonumber \\
    &\geq& |\cos\psi_k\cos\beta_k-\sin\psi_k\sin\gamma_k|.
\end{eqnarray}
Since $0 \leq \sin\gamma_k \leq \sin\beta_k$, $\cos(\psi_k+\beta_k) \leq \cos\psi_k\cos\beta_k-\sin\psi_k\sin\gamma_k \leq \cos\psi_k\cos\beta_k$. On the other hand, since $\cos\theta \leq \cos\psi_k < 1$ and $\cos(k\theta) \leq \cos\beta_k < 1$, we have $\cos(\psi_k+\beta_k) \geq \cos((k+1)\theta)$.

If $(k+1)\theta \geq \frac{\pi}{2}$, $|\cos\psi_k\cos\beta_k-\sin\psi_k\sin\gamma_k| \geq 0$ with possibility to reach equality. If $t_k = \f(\E(c_k),b_k) = 0$, it is done. Then $N_{\min} \geq \biggl\lceil\frac{\pi}{2\arccos(\f_1(\E,\I))}\biggr\rceil$. Otherwise $(k+1)\theta < \frac{\pi}{2}$, $t_k \geq |\cos\psi_k\cos\beta_k-\sin\psi_k\sin\gamma_k| \geq \cos((k+1)\theta) > 0$. Further queries are inevitably needed. Therefore we prove that the lower bounds also hold for the case of $k+1$.

By induction, we prove that the above lower bounds hold for every integer $k \geq 1$. Therefore
\begin{equation}
    N_{\min} \geq \biggl\lceil\frac{\pi}{2\arccos(\f_1(\E,\I))}\biggr\rceil.
\end{equation}
\hfill $\square$

Under some circumstances, the lower bound is reachable. For example, if $\exists \ket{b}$ such that $E_i\ket{b} = \lambda_i\ket{c}(i = 1 \ldots m, 2 \leq m \leq d^2-1)$ and $|\ip{c}{b}| = \f_1(\E,\I)$, using the construction method similar to that in Theorem \ref{seq2dim}, we can obtain the optimal scheme reaching the lower bound. In Theorem \ref{seqddim}, when $m=1$, the lower bound matches the conclusion in \cite{DFY07} perfectly.

\section{General Optimal Perfect Distinguishability between Quantum Operations}
\noindent
We can generalize the above sequential results to the most general discrimination situations, where auxiliary systems and entanglement are acceptable. Since Lemma \ref{upperbound} works for arbitrary dimensions, we have
\begin{equation}
    \f_q^{ea}(\E,\I) \leq \frac{|\sin(\alpha_0'-\alpha)|}{\sin\alpha_0'}\cos\theta'.
\end{equation}
Note that here $\f_1^{ea}(\E,\I)=\cos\theta'(0 < \theta' < \frac{\pi}{2})$ and $\cos\alpha_0' = \max\{|\ip{b}{b'}|: \f((\I^R \otimes \E^Q)(b),(\I^R \otimes \I^Q)(b)) = \cos\theta', \op{b}{b'} \in {\rm span}^\perp\{I \otimes E_i\}_{i = 1 \ldots m}\}(0 \leq \alpha_0' \leq \frac{\pi}{2})$. By \cite{DFY09}, if perfectly distinguishable, it must be $\f_1^{ea}(\E,\I)=\cos\theta'<1$, or $\theta'>0$. If $\theta' = \frac{\pi}{2}$, obviously one query is done. Otherwise, $0 < \theta' < \frac{\pi}{2}$. According to the proof of Lemma \ref{upperbound}, we know that $\alpha_0'$ exists and $0 < \alpha_0' < \frac{\pi}{2}$, hence the above inequality holds. Consequently, under the premise that a unitary and a quantum operation are perfectly distinguishable, we can obtain an upper bound for optimal query time:
\begin{equation}
    N_{\min} \leq \biggl\lceil\frac{\ln \cos\alpha_0'}{\ln \cos\theta'}\biggr\rceil+1.
\end{equation}
Interestingly, $\cos\theta'$ and $\cos\alpha_0'$ correspond to $q_1$ and $q_{\max}$ in \cite{DFY09} respectively and the two upper bounds coincide. While \cite{DFY09} arrives the conclusion based on entanglement, we are independent of auxiliary systems.

Continuing the ideas, for arbitrary $d$-dimensional quantum operations $\E_0$ and $\E_1$, we have the following theorem:
\begin{theorem}\label{general}\Upshape
Suppose $q = \cos\alpha(0 \leq \alpha \leq \frac{\pi}{2})$, $\f_1(\E_0,\I)=\cos\theta_0(0 \leq \theta_0 \leq \frac{\pi}{2})$ and $\f_1(\E_1,\I)=\cos\theta_1(0 \leq \theta_1 \leq \frac{\pi}{2})$. If $\alpha+\theta_0 \geq \frac{\pi}{2}$ or $\alpha+\theta_1 \geq \frac{\pi}{2}$ or $\alpha+\theta_0+\theta_1 \geq \frac{\pi}{2}$, $\f_q(\E_0,\E_1) \geq 0$. Otherwise, $\f_q(\E_0,\E_1) \geq \cos(\alpha+\theta_0+\theta_1)$.
\end{theorem}

{\bf Proof:} Consider arbitrary pair of input states $\{\ket{b_0},\ket{b_1}\}$ such that $|\ip{b_0}{b_1}|=q=\cos\alpha$ for $\E_0$ and $\E_1$ respectively. Assume $\ket{b_0'} \in {\rm supp}(\E_0(b_0))$ and $\ket{b_1'} \in {\rm supp}(\E_1(b_1))$ satisfying $|\ip{b_0'}{b_0}| = \f(\E_0(b_0),b_0) = \cos\beta_0$ and $|\ip{b_1'}{b_1}| = \f(\E_1(b_1),b_1) = \cos\beta_1$. Hence $0 \leq \beta_0 \leq \theta_0$ and $0 \leq \beta_1 \leq \theta_1$. Same as the proof of Theorem \ref{seqddim}, we can represent $\ket{b'_0}$ and $\ket{b'_1}$ as
\begin{eqnarray}
\ket{b'_0} &=& \cos\beta_0\ket{b_0} + \sin\beta_0\ket{b_0^\perp}, \\
\ket{b'_1} &=& \cos\beta_1\ket{b_1} + \sin\beta_1\ket{b_1^\perp},
\end{eqnarray}
$\ip{b_0}{b_0^\perp}=\ip{b_1}{b_1^\perp}=0$.

Let $t=\f(\E_0(b_0),\E_1(b_1))$,
\begin{eqnarray}
	t &\geq& |\ip{b'_0}{b'_1}|\nonumber \\
	&=&|\cos\beta_0\cos\beta_1\ip{b_0}{b_1} + \cos\beta_0\sin\beta_1\ip{b_0}{b_1^\perp}\nonumber \\
    &\quad& + \sin\beta_0\cos\beta_1\ip{b_0^\perp}{b_1} + \sin\beta_0\sin\beta_1\ip{b_0^\perp}{b_1^\perp}|\nonumber \\
	&\geq& |a-b|,
\end{eqnarray}
where
\begin{eqnarray}
a &=& |\cos\beta_0\cos\beta_1\ip{b_0}{b_1} + \cos\beta_0\sin\beta_1\ip{b_0}{b_1^\perp}|, \\
b &=& |\sin\beta_0\cos\beta_1\ip{b_0^\perp}{b_1} + \sin\beta_0\sin\beta_1\ip{b_0^\perp}{b_1^\perp}|.
\end{eqnarray}
Since $|\ip{b_0}{b_1}| = \cos\alpha$ and $\ip{b_0}{b_1}\ip{b_1}{b_0}+\ip{b_0}{b_1^\perp}\ip{b_1^\perp}{b_0} \leq 1$, $0 \leq |\ip{b_0}{b_1^\perp}| \leq \sin\alpha$. Similarly, $0 \leq |\ip{b_0^\perp}{b_1}|=\sin\zeta \leq \sin\alpha(0 \leq \zeta \leq \alpha)$ and $0 \leq |\ip{b_0^\perp}{b_1^\perp}| \leq \sqrt{1-|\ip{b_0^\perp}{b_1}|^2} = \cos\zeta$.

Now we can obtain upper bounds and the lower bounds of $a$ and $b$ respectively:
\begin{enumerate}
\item $a$'s upper bound:
\begin{eqnarray}
a &\leq& \cos\beta_0(\cos\beta_1|\ip{b_0}{b_1}| + \sin\beta_1|\ip{b_0}{b_1^\perp}|)\nonumber \\
&\leq& \cos\beta_0\cos(\beta_1-\alpha).
\end{eqnarray}

\item $a$'s lower bound:
\begin{eqnarray}
a &\geq& \cos\beta_0\big|\cos\beta_1|\ip{b_0}{b_1}|-\sin\beta_1|\ip{b_0}{b_1^\perp}|\big|\nonumber \\
&\geq& \cos\beta_0\big|\cos\beta_1\cos\alpha-\sin\beta_1|\ip{b_0}{b_1^\perp}|\big|.
\end{eqnarray}
If $\beta_1+\alpha < \frac{\pi}{2}$, $a \geq \cos\beta_0|\cos\beta_1\cos\alpha-\sin\beta_1\sin\alpha| = \cos\beta_0\cos(\beta_1+\alpha)$. Otherwise $\beta_1+\alpha \geq \frac{\pi}{2}$, $a \geq 0$.

\item $b$'s upper bound:
\begin{eqnarray}
b &\leq& \sin\beta_0(\cos\beta_1|\ip{b_0^\perp}{b_1}| + \sin\beta_1|\ip{b_0^\perp}{b_1^\perp}|)\nonumber \\
&\leq& \sin\beta_0(\cos\beta_1\sin\zeta + \sin\beta_1\cos\zeta)\nonumber \\
&=& \sin\beta_0\sin(\beta_1+\zeta).
\end{eqnarray}
If $\beta_1 + \alpha < \frac{\pi}{2}$, $b \leq \sin\beta_0\sin(\beta_1+\zeta) \leq \sin\beta_0\sin(\beta_1+\alpha)$. Otherwise $\beta_1 + \alpha \geq \frac{\pi}{2}$, $b \leq \sin\beta_0$.

\item $b$'s lower bound:
\begin{eqnarray}
b &\geq& \sin\beta_0\big|\cos\beta_1|\ip{b_0^\perp}{b_1}| - \sin\beta_1|\ip{b_0^\perp}{b_1^\perp}|\big|\nonumber \\
&=& \sin\beta_0\big|\cos\beta_1\sin\zeta - \sin\beta_1|\ip{b_0^\perp}{b_1^\perp}|\big|\nonumber \\
&\geq& 0.
\end{eqnarray}
\end{enumerate}

Combine the above results, if $\beta_1+\alpha < \frac{\pi}{2}$, $\cos\beta_0\cos(\beta_1+\alpha) \leq a \leq \cos\beta_0\cos(\beta_1-\alpha)$, $0 \leq b \leq \sin\beta_0\sin(\beta_1+\alpha)$.
\begin{eqnarray}
\cos\beta_0\cos(\beta_1+\alpha)&-&\sin\beta_0\sin(\beta_1+\alpha) = \cos(\alpha+\beta_0+\beta_1)\nonumber \\
&\leq& a-b \leq \cos\beta_0\cos(\beta_1-\alpha).
\end{eqnarray}
When $\alpha+\beta_0+\beta_1 < \frac{\pi}{2}$,
\begin{equation}
t \geq |a-b| \geq \cos(\alpha+\beta_0+\beta_1).
\end{equation}
When $\alpha+\beta_0+\beta_1 \geq \frac{\pi}{2}$,
\begin{equation}
t \geq |a-b| \geq 0.
\end{equation}

Otherwise $\beta_1+\alpha \geq \frac{\pi}{2}$, $0 \leq a \leq \cos\beta_0\cos(\beta_1-\alpha)$, $0 \leq b \leq \sin\beta_0$.
\begin{equation}
-\sin\beta_0 \leq a-b \leq \cos\beta_0\cos(\beta_1-\alpha).
\end{equation}
Hence
\begin{equation}
t \geq |a-b| \geq 0.
\end{equation}

Substitute $\beta_1$ with $\beta_0$, by symmetry, if $\beta_0 + \alpha < \frac{\pi}{2}$, when $\alpha+\beta_0+\beta_1 < \frac{\pi}{2}$,
\begin{equation}
	t \geq |a-b| \geq \cos(\alpha+\beta_0+\beta_1).
\end{equation}
When $\alpha+\beta_0+\beta_1 \geq \frac{\pi}{2}$,
\begin{equation}
	t \geq |a-b| \geq 0.
\end{equation}
Otherwise $\beta_0 + \alpha \geq \frac{\pi}{2}$,
\begin{equation}
	t \geq |a-b| \geq 0.
\end{equation}

Finally, by the randomicity of $\{\ket{b_0},\ket{b_1}\}$ and $0 \leq \beta_0 \leq \theta_0, 0 \leq \beta_1 \leq \theta_1$, the conclusions hold.
\hfill $\square$

Substitute $\E_0$ and $\E_1$ in Theorem \ref{general} with $\I^R \otimes \E^Q$ and $\I^R \otimes \I^Q$ respectively, we have
\begin{corollary}
If a unitary and a quantum operation are perfectly distinguishable, the optimal query time $N_{\min}$ satisfies:
\begin{equation}
    N_{\min} \geq \biggl\lceil\frac{\pi}{2\arccos(\f_1^{ea}(\E,\I))} \biggr\rceil.
\end{equation}
\end{corollary}

\begin{corollary}
If quantum operations $\E_0$ and $\E_1$ are sequentially perfectly distinguishable, the optimal query time $N_{\min}$ satisfies:
\begin{equation}
    N_{\min} \geq \biggl\lceil\frac{\pi}{2(\arccos(\f_1(\E_0,\I))+\arccos(\f_1(\E_1,\I)))} \biggr\rceil.
\end{equation}
\end{corollary}

Finally, substitute $\E_0$ and $\E_1$ in Theorem \ref{general} with $\I^R \otimes \E_0^Q$ and $\I^R \otimes \E_1^Q$ respectively, we have
\begin{corollary}
If quantum operations $\E_0$ and $\E_1$ are perfectly distinguishable, the optimal query time $N_{\min}$ satisfies:
\begin{equation}
    N_{\min} \geq \biggl\lceil\frac{\pi}{2(\arccos(\f_1^{ea}(\E_0,\I))+\arccos(\f_1^{ea}(\E_1,\I)))} \biggr\rceil.
\end{equation}
\end{corollary}

\section{Future Directions}
\noindent
There are many research directions regarding to optimal discrimination. It is natural to explore optimal parallel perfect distinguishability between a unitary and a quantum operation. The case between two unitaries has been solved in \cite{AC01}. As a general upper bound and lower bound have been obtained, it would be interesting to consider whether we can obtain a tighter bound in parallel or design a parallel scheme to reach the lower bound. What's more, the role of entanglement in parallel discrimination remains open. According to Theorem 2 in \cite{DFY09}, if we can figure out the series $\{q_k\}_{k \geq 1}$, we can obtain accurate optimal query time. To do this, we may need the tool of semi-definite programming, which may be another interesting direction to reach a better characterization over optimal query time under general circumstances.

\section*{Acknowledgment}
\noindent
We are grateful to Nengkun Yu for helpful discussions at the early stage of this work.

\end{document}